\documentclass[notitlepage,aps,twocolumn,floatfix,nofootinbib]{revtex4-1}

\usepackage{graphicx,amsmath,amssymb,verbatim,color}
\usepackage{booktabs}
\usepackage{fancyhdr, fancybox, empheq} 
\usepackage{color}
\usepackage{url}
\usepackage{subcaption}

\usepackage[colorlinks=true,citecolor=blue,linkcolor=blue,urlcolor=blue, backref=false,pdfborder={0 0 0}]{hyperref}

\newcommand{\physrep}{Phys.~Rep.}
\newcommand{\aap}{Astron.~Astrophys.}
\newcommand{\jcap}{JCAP}
\newcommand{\apjl}{Astrophys. J. Lett.}
\newcommand{\mnras}{Mon. Not. Roy. Ast. Soc.}

\newcommand{\bl}{\boldsymbol{l}}
\newcommand{\bll}{\boldsymbol{L}}

\newcommand{\vsp}{\vphantom{\Big[}\\}
\newcommand{\dotfac}[1]{({\bll} \cdot {\bl}_{#1})}

\newcommand{\intl}[1]{\int {d^2 l_{#1} \over (2\pi)^2}}

\newcommand{\yah}[1]{{\textcolor{blue}{#1}}}
\newcommand{\beq}{\begin{equation}}
\newcommand{\eeq}{\end{equation}}
\newcommand{\barr}{\begin{eqnarray}}
\newcommand{\earr}{\end{eqnarray}}
\newcommand{\bs}{\boldsymbol}

\newcommand{\intL}{\int_{\substack{\bl_1 + \bl_2 \\ =\bll }}}
\newcommand{\intLp}{\int_{\substack{\bl'_1 + \bl'_2 \\ =\bll' }}}

\begin{document}

\title{Quadratic estimators for CMB weak lensing}
\author{Abhishek S. Maniyar}
\email{abhishek.maniyar@nyu.edu}
\affiliation{Center for Cosmology and Particle Physics, Department of Physics, New York University, New York, NY 10003, USA}
\author{Yacine Ali-Ha\"imoud}
\affiliation{Center for Cosmology and Particle Physics, Department of Physics, New York University, New York, NY 10003, USA}
\author{Julien Carron}
\affiliation{Universit\'e de Gen\`eve, D\' epartement de Physique Th\'eorique et CAP, 24 Quai Ansermet, CH-1211 Gen\`eve 4, Switzerland}
\author{Antony Lewis}
\affiliation{Department of Physics and Astronomy, University of Sussex, Brighton BN1 9QH, UK}
\author{Mathew S. Madhavacheril}
\affiliation{Perimeter Institute for Theoretical Physics, Waterloo, ON, Canada N2L 2Y5}


\begin{abstract}
In recent years, weak lensing of the cosmic microwave background (CMB) has emerged as a powerful tool to probe fundamental physics, such as neutrino masses, primordial non-Gaussianity, dark energy, and modified gravity. The prime target of CMB lensing surveys is the lensing potential, which is reconstructed from the observed CMB temperature $T$ and polarization $E$ and $B$ fields. Until very recently, this reconstruction has been performed with quadratic estimators (QEs), which, although known to be suboptimal for high-sensitivity experiments, are numerically efficient, and useful to make forecasts and cross-check the results of more sophisticated likelihood-based methods. It is expected that ongoing and near-future CMB experiments such as AdvACT, SPT-3G and the Simons Observatory (SO), will also rely on QEs. In this work, we review different QEs, and clarify and quantify their differences. In particular, we show that the Hu-Okamoto (HO02) estimator is not the absolute optimal lensing estimator that can be constructed out of quadratic combinations of $T, E$ and $B$ fields. Instead, we derive the global-minimum-variance (GMV) lensing quadratic estimator. Although this estimator can be found elsewhere in the literature, it was erroneously described as equivalent to the HO02 estimator, and has never been used in real data analyses. Here, we show explicitly that the HO02 estimator is suboptimal to the GMV estimator, with a reconstruction noise larger by up to $\sim 9\%$ for a SO-like experiment. We further show that the QE used in the Planck, and recent SPT lensing analysis is suboptimal to both the HO02 and GMV estimator, and would have a reconstruction noise up to $\sim 11\%$ larger than that of the GMV estimator for a SO-like experiment. In addition to clarifying differences between different QEs, this work should thus provide motivation to implement the GMV estimator in future lensing analyses relying on QEs.
\end{abstract}

\maketitle

\section{Introduction} \label{sec:intro}

Weak gravitational lensing of the cosmic microwave background (CMB) arises from the deflection of CMB photons as they travel to us from the last scattering surface, through the inhomogeneous Universe \cite{Blanchard_87}; see e.g.~Ref.~\cite{Lewis_06} for a review. The deflection angle is proportional to the gradient of the lensing potential $\phi$, which is determined by the projected mass distribution along the line of sight. Reconstructing $\phi$ is therefore a powerful cosmological tool, as it gives direct access to the projected distribution of the \emph{total} matter -- baryonic and dark -- without relying on biased tracers \cite{Seljak_99}. Among other applications, the power spectrum of the lensing potential and its cross-correlation with other tracers of large-scale structure are a sensitive probe of the growth of matter fluctuations, primordial non-Gaussianity, neutrino masses, dark energy, and modified gravity \cite{Lewis_06, Allison_15, Schmittfull_18}. CMB lensing has been successfully measured by ACT, SPT, \textit{Planck}, BICEP and POLARBEAR \cite{Das_11, Sherwin_17, van_12, Planck_lensing_16, Planck_lensing_18, Omori_17, Story_15, Wu_19, Millea_20b, BICEP_lensing, Ade_14, Adachi_20}. Current and upcoming wide-field CMB experiments such as  AdvACT \cite{Henderson_16}, SPT-3G \cite{Benson_14} and the Simons Observatory (SO) \cite{Ade_19} will measure the lensing potential with even higher signal-to-noise ratio. Looking ahead, next-generation ``Stage-4" instrumental concepts with unprecedented depth and angular resolution are currently under development \cite{Abazajian_16}, with CMB lensing as one of their main science goals \cite{S4_19}.

One of the main signatures of weak lensing is the induced correlations between unequal Fourier modes of the CMB temperature and polarization fields. It is therefore natural to seek to estimate $\phi$ out of linear combinations of terms quadratic in different modes of the observed fields \cite{Zaldarriaga_99, Hu_01}; and indeed, almost all CMB lensing analyses thus far have relied on such quadratic estimators. For the next-generation Stage-4-like CMB experiments (CMBS4), quadratic estimators are known to be suboptimal, especially for polarization \cite{Hirata_03}. More elaborate algorithms are being developed, such as the gradient-inversion method \cite{Hadzhiyska_19}, or likelihood-based methods \cite{Hirata_03a, Hirata_03, Carron_17, Millea_20}. Meanwhile, quadratic estimators remain the workhorse tool for current and near-future CMB experiments like AdvACT, SPT-3G, and SO. They have the advantages of being very simple to implement and computationally efficient, and will serve as useful cross-checks even when more accurate and computationally demanding methods are employed with future data. 

The main goal of this paper is to clarify and quantify the differences between several quadratic estimators commonly used for CMB lensing reconstruction. Our most important point is that the well-known Hu and Okamoto \cite{Hu_02} (hereafter, HO02) estimator is \emph{not} the optimal quadratic estimator that can be constructed from temperature and polarization maps, even if generalized to the full sky, and even when using non-perturbative response functions \cite{Lewis_11, Fabbian_19}. Instead, we derive the global-minimum-variance (hereafter GMV) quadratic estimator built out of all possible quadratic combination of $T, E$ and $B$. The GMV estimator was in fact first derived in Hirata and Seljak \cite{Hirata_03}, as the weak-signal limit of their likelihood-based method. Nevertheless, it was stated there and in subsequent works that this estimator is equivalent to that of HO02. We explicitly show that this is not the case, and that the reconstruction noise of the GMV estimator can be up to $\sim 9\%$ lower than that of the HO02 estimator on large angular scales. We also generalize it to be accurately unbiased accounting for higher-order lensing effects.
Furthermore, we show that the quadratic estimator used in the Planck collaboration \cite{Planck_lensing_16, Planck_lensing_18} and SPT collaboration \cite{Wu_19} lensing analyses, obtained by neglecting $C^{TE}_\ell$ in the inverse filter matrix, is suboptimal to \emph{both} the GMV and HO02 estimators. For a SO-like experiment, this suboptimal estimator is up to $\sim 11\%$ noisier that the GMV estimator. This may motivate implementing the GMV estimator in future analyses, despite the possible added complexity of jointly filtering temperature and polarization maps. 

The remainder of this paper is organized as follows. After introducing our notation and convention in Sec.~\ref{sec:notation}, we review the HO02 estimator and its close cousin, the Okamoto-Hu \cite{Hu_03} (hereafter OH03) estimator in Sec.~\ref{sec:Hu-Okamoto}. We then derive the GMV estimator in Sec.~\ref{sec:GMV} and explicitly show how it differs from the HO02 estimator. We describe the suboptimal lensing estimator of Ref.~\cite{Planck_lensing_18} in Sec.~\ref{sec:SQE}. Finally, we compare estimators in Sec.~\ref{sec:comparison} for different instrumental setups, and conclude in Sec.~\ref{sec:conclusion}.

\section{Notation and conventions} \label{sec:notation}

We denote by capital letters $X, Y = T, E, B$ the observed (lensed and noisy) CMB temperature and polarization fields, and by $\phi$ the projected lensing potential. Throughout we work in the flat-sky approximation; we denote two-dimensional Fourier wavenumbers by $\bl$ for CMB fields and $\bll$ for the lensing potential. 

The power spectra of the observed temperature and polarization fields are defined as
\barr
\langle X(\bl) Y(\bl') \rangle = (2\pi)^2 \delta(\bl+\bl') C_l^{XY} \, ,
\label{eq:cmbps}
\earr
where $C_l^{XY}$ is the total cross-power spectrum of the lensed fields, including detector noise added in quadrature (for $X = Y$). It can also include contributions from other sources of variance such as residual foreground contamination. In this expression the angular brackets denote taking ensemble averages over the primordial CMB, detector noise, along with the underlying large scale structure. 

Gravitational lensing affects the auto- and cross-power spectra of CMB fields, and moreover produces correlations between non-opposite $\bl$ modes, proportional to the projected lensing potential. The response of the non-opposite correlations to lensing can be quantified by non-perturbative response functions $f_{XY}$ defined by
  \beq
  \left\langle \frac{\delta}{\delta \phi(\bll)}\left(  X(\bl)Y(\bl')\right)\right\rangle
  = \delta(\bl+\bl'-\bll) f_{XY}(\bl,\bl').
\label{eq:cmbavg}
  \eeq


The coupling coefficients $f_{XY}$ appearing in Eq.~\eqref{eq:cmbavg} are given explicitly in Table \ref{tab:falpha}. They depend on the lensed gradient spectra $\widetilde{C}_l^{X \nabla Y}$ defined in Refs.~\cite{Lewis_11, Fabbian_19}, which generalize the unlensed spectra used in the original work of HO02 so that the response function for each lensing mode includes the important higher-order effect of other lensing modes. The BB term has negligible contribution to the signal-to-noise ratio of the reconstructed $\phi$ field and thus we neglect it in our analysis. Note that different foregrounds can also contribute to off-diagonal correlations \citep[e.g.][]{Ferraro_18, Schaan_19}, but we do not include them in this work.

\begin{table}
\begin{center}
\begin{tabular}{c|c}
$\alpha$ 	 & $f_\alpha({\bl_1,\bl_2})$ \vsp
\hline
$TT$ 	& $\widetilde{C}_{l_1}^{T \nabla T}  \dotfac{1} +  \widetilde{C}_{l_2}^{T \nabla T}  \dotfac{2}$\vsp
$TE$	& $\widetilde{C}_{l_1}^{T \nabla E}\cos 2\varphi_{\bl_1\bl_2} \dotfac{1} +  \widetilde{C}_{l_2}^{T \nabla E}\dotfac{2}$\vsp
$EE$	& $[\widetilde{C}_{l_1}^{E \nabla E} \dotfac{1} +\widetilde{C}_{l_2}^{E \nabla E} \dotfac{2}]\cos 2 \varphi_{\bl_1\bl_2} $ \vsp
$TB$	& $\widetilde{C}_{l_1}^{T \nabla E}\sin 2\varphi_{\bl_1\bl_2} \dotfac{1}$\vsp
$EB$	& $[\widetilde{C}_{l_1}^{E \nabla E} \dotfac{1} + \widetilde{C}_{l_2}^{B \nabla B} \dotfac{2}]\sin 2 \varphi_{\bl_1\bl_2} $ \vsp
\end{tabular}
\end{center}
\caption{CMB lensing correlation coefficients. $\varphi_{\bl_1 \bl_2}$ is the angle between $\bl_1$ and $\bl_2$. The quantity $\widetilde{C}^{X \nabla Y}$ is the lensed gradient spectrum, defined in Refs.~\cite{Lewis_11, Fabbian_19}. Note that we do not include curl-like terms $\widetilde{C}_l^{T P_{\bot}}, \widetilde{C}_l^{PP_{\bot}}$, which are always subdominant \cite{Fabbian_19}.}\label{tab:falpha}
\end{table}

In the remainder of this work we will describe different estimators $\hat{\phi}_\alpha$ for the lensing potential. All these estimators are required to be unbiased, i.e.~such that $\langle \hat{\phi}_{\alpha} \rangle = \phi$. They are however noisy, and we define their variance (or reconstruction noise) $N_\alpha(L)$ through  
\beq
\langle (\hat{\phi}_{\alpha}-\phi)(\bll) (\hat{\phi}_{\alpha}-\phi)(\bll') \rangle = (2\pi)^2 \delta_{\rm D}(\bll+\bll') N_\alpha(L). \label{eq:est-noise}
\eeq
Here, for optimizing the signal to noise, we only consider the primary Gaussian disconnected contractions of the lensed fields, $N_\alpha ^{(0)}(L)$; in Appendix.~\ref{app:NOne} we give an  explicit form for the  $N_\alpha^{(1)}(L)$ contractions \cite{Kesden_03} that should also be included in any full data likelihood analysis. The superscript values 0 and 1 represent the order to which the variance $N_\alpha(L)$ explicitly depends on $C^{\phi \phi}_L$.

We will often deal with convolutions in Fourier space, and for brevity, introduce the compact notation
\beq
\intL ... \equiv \iint \frac{d^2 l_1 d^2 l_2}{(2 \pi)^2} \delta_{\rm D}(\bl_1 + \bl_2 - \bll) ...
\eeq

We define the Fourier transform of a configuration-space function $A(\boldsymbol{\hat{n}})$ as 
\beq
\mathcal{F}[A(\boldsymbol{\hat{n}})](\bl) \equiv \int d^2 \boldsymbol{\hat{n}} ~A(\boldsymbol{\hat{n}})  \textrm{e}^{-i \bl \cdot \boldsymbol{\hat{n}}}, 
\eeq
and the inverse-Fourier transform of a harmonic-space function $B(\bl)$ as
\beq
\mathcal{F}^{-1}[B(\bl)](\boldsymbol{\hat{n}}) \equiv \int \frac{d^2 \bl}{(2 \pi)^2} B(\bl) \textrm{e}^{i \bl \cdot \boldsymbol{\hat{n}}}.
\eeq

\section{Hu and Okamoto estimators} \label{sec:Hu-Okamoto}

We now briefly rederive the HO02 and OH03 quadratic estimators for the lensing potential, setting the stage for our subsequent derivation of the global-minimum-variance estimator. 

The approach of HO02 consists in constructing the single-pair estimators $\hat{\phi}_\alpha({\bll})$ separately for each pair $\alpha = TT, TE, EE, TB, EB$, and then combining them together to form their minimum variance estimator. 

OH03 moreover derive efficient full-sky single-pair estimators in configuration space. They are identical to the HO02 estimators, except for the $TE$ estimator, which is slightly sub-optimal. We give explicit expressions for these estimators in the flat-sky limit in Section \ref{sec:OH}. Here again, the final OH03 estimator is obtained by combining these single-pair estimators. 
Note that HO02 and OH03 used unlensed spectra in the response functions rather than the lensed gradient spectra. We still refer to the estimators constructed with the lensed gradient spectra as the HO02 and OH03 estimators, given that the procedure is identical.

\subsection{Single-pair minimum-variance quadratic estimators in harmonic space}

We start by constructing quadratic estimators out of a single pair $XY$, of the form
\beq
\hat{\phi}_{XY}(\bll) \equiv \intL X(\bl_1) Y(\bl_2) F_{XY}(\bl_1, \bl_2). \label{eq:estimator}
\eeq
For the estimator to be unbiased, the weights $F_{XY}$ must satisfy the constraint
\beq
\intL f_{XY}(\bl_1, \bl_2) F_{XY}(\bl_1, \bl_2) = 1. \label{eq:unbiased}
\eeq
The noise of this estimator (defined as in Eq.~\eqref{eq:est-noise}) is 
\barr
N_{XY}(L) = \intL F_{XY}(\bl_1, \bl_2) \Big( F_{XY}(\bl_1, \bl_2) C_{l_1}^{XX} C_{l_2}^{YY} \nonumber\\
+ F_{XY}(\bl_2, \bl_1) C_{l_1}^{XY} C_{l_2}^{XY} \Big).~~~ \label{eq:variance}
\earr

\subsubsection{All pairs except $TE$}

For all pairs except $TE$, either $X = Y$ or $C_l^{XY}$ = 0. As a consequence, the variance of the estimator takes the form
\barr
N_{XY}(L) = (1 + \delta_{XY}) \intL  C_{l_1}^{XX} C_{l_2}^{YY} [F_{XY}(\bl_1, \bl_2)]^2.~~~~ \label{eq:NXY-not-TE}
\earr
Minimizing this variance under the constraint \eqref{eq:unbiased} results in the following coefficients
\barr
\label{eq:F_XX}
F_{XY}(\bl_1, \bl_2) &=& \lambda_{XY}(L) ~\frac{f_{XY}(\bl_1, \bl_2)}{(1 + \delta_{XY}) C_{l_1}^{XX} C_{l_2}^{YY}}, \label{eq:FXY-not-TE}\\
\lambda_{XY}(L) &\equiv& \left[\intL \frac{[f_{XY}(\bl_1, \bl_2)]^2}{(1 + \delta_{XY}) C_{l_1}^{XX} C_{l_2}^{YY}}\right]^{-1}.
\earr
Inserting back into Eq.~\eqref{eq:NXY-not-TE}, we find the corresponding minimum variance $N_{XY}(L) = \lambda_{XY}(L)$.

\subsubsection{Special case of $XY = TE$}

We may decompose $F_{TE}$ into a symmetric and antisymmetric piece:
\barr
F_{TE}(\bl_1, \bl_2) &=& F_{TE}^{+}(\bl_1, \bl_2) + F_{TE}^{-}(\bl_1, \bl_2), \\
F_{TE}^{\pm}(\bl_1, \bl_2) &\equiv& \frac12 \left(F_{TE}(\bl_1, \bl_2) \pm F_{TE}(\bl_2, \bl_1) \right). \label{eq:Fpm}
\earr
For each pair $(\bl_1, \bl_2)$, we further define the 2-dimensional vector
\beq
\bs{F}(\bl_1, \bl_2) \equiv \left( F_{TE}^{+}(\bl_1, \bl_2) , F_{TE}^{-}(\bl_1, \bl_2)\right).
\eeq
After some algebra, and only keeping even functions of $(\bl_1, \bl_2)$ in the integral, Eq.~\eqref{eq:variance} can be rewritten as
\beq
N_{TE}(L) = \intL \bs{F}(\bl_1, \bl_2) \cdot \bs{M}(l_1, l_2) \cdot \bs{F}(\bl_1, \bl_2),  \label{eq:var_XY}
\eeq
where for each pair $(l_1, l_2)$, the 2 by 2 matrix $\bs{M}(l_1, l_2)$ is given by
\barr
&&\bs{M}(l_1, l_2) = \nonumber\\
&&\begin{pmatrix}C_{(l_1}^{TT} C_{l_2)}^{EE} + C_{l_1}^{TE} C_{l_2}^{TE} & C_{[l_1}^{TT} C_{l_2]}^{EE}\\
C_{[l_1}^{TT} C_{l_2]}^{EE} & C_{(l_1}^{TT} C_{l_2)}^{EE} - C_{l_1}^{TE} C_{l_2}^{TE}\end{pmatrix},~~~ 
\earr
where $A_{(l_1 l_2)} \equiv (A_{l_1 l_2} + A_{l_2 l_1})/2$ and $A_{[l_1 l_2]} \equiv (A_{l_1 l_2} - A_{l_2 l_1})/2$ are the symmetric and antisymmetric parts of $A_{l_1 l_2}$.

Similarly, we may define the symmetric and antisymmetric parts of the correlation coefficients $f_{TE}^{\pm}(\bl_1, \bl_2)$ and the two-dimensional vector $\bs{f} = (f_{TE}^+, f_{TE}^-)$, for each pair $(\bl_1, \bl_2)$, and rewrite the constraint \eqref{eq:unbiased} as
\barr
\intL \bs{F}(\bl_1, \bl_2) \cdot \bs{f}(\bl_1, \bl_2)  = 1.\label{eq:constr}
\earr
Minimizing the variance \eqref{eq:var_XY} under this constraint leads to the solution
\beq
\bs{F}(\bl_1, \bl_2) = \lambda(L) ~\bs{M}^{-1} (\bl_1, \bl_2) \cdot \bs{f}(\bl_1, \bl_2), \label{eq:F-sol}
\eeq
where the Lagrange multiplier $\lambda$ is obtained from the constraint \eqref{eq:constr}:
\beq
\lambda(L) = \left(\intL \bs{f}(\bl_1, \bl_2) \cdot \bs{M}^{-1} (l_1, l_2) \cdot \bs{f}(\bl_1, \bl_2) \right)^{-1}. 
\eeq
The 2 by 2 matrix $\bs{M}(l_1, l_2)$ is easily invertible, and after re-expressing Eq.~\eqref{eq:F-sol} in terms of the original $F_{XY}(\bl_1, \bl_2)$ and $f_{TE}(\bl_1, \bl_2)$, one recovers the HO02 optimal weights for $TE$, namely, with our notation, 
\begin{widetext}
\barr
\label{eq:F_XY}
F_{TE}(\bl_1, \bl_2) &=& \lambda_{TE}(L)\frac{C_{l_1}^{EE} C_{l_2}^{TT} f_{TE}(\bl_1, \bl_2) - C_{l_1}^{TE} C_{l_2}^{TE}  f_{TE}(\bl_2, \bl_1)}{C_{l_1}^{TT} C_{l_2}^{EE}C_{l_1}^{EE} C_{l_2}^{TT} - \left(C_{l_1}^{TE} C_{l_2}^{TE}\right)^2},  \\
\lambda_{TE}(L) &\equiv& \Bigg[\intL f_{TE}(\bl_1, \bl_2) \frac{C_{l_1}^{EE} C_{l_2}^{TT} f_{TE}(\bl_1,\bl_2) - C_{l_1}^{TE} C_{l_2}^{TE} f_{TE}(\bl_2,\bl_1)}{C_{l_1}^{TT} C_{l_2}^{EE}C_{l_1}^{EE} C_{l_2}^{TT} - \left(C_{l_1}^{TE} C_{l_2}^{TE}\right)^2} \Bigg]^{-1} \, .
\earr
\end{widetext}
Inserting Eq.~\eqref{eq:F-sol} into Eq.~\eqref{eq:var_XY}, we see that the noise of the minimum-variance estimator is just $N_{TE}(L) = \lambda_{TE}(L)$.

\subsection{Single-pair efficient configuration-space estimators} \label{sec:OH}

\subsubsection{All pairs except $TE$}

The response coefficients $f_{XY}(\bl_1, \bl_2)$ can all be written as linear combinations of products of functions of $\bl_1$ with functions of $\bl_2$, with coefficients depending on $L$. From Eq.~\eqref{eq:FXY-not-TE}, we see that this property transfers to the optimal weights $F_{XY}$ for all pairs except $TE$. As a consequence, all single-pair estimators except $TE$ can be written as sums of convolutions of functions of $\bl_1$ with functions of $\bl_2$. This implies that they can be written as a sum of products of functions of configuration space -- they are ``separable" in configuration space. This allows to use Fast Fourier Transforms (FFTs) (or fast harmonic transforms for full-sky expressions \cite{Hu_03}) to compute them efficiently.

Similar to OH03, we define the following bilinear operator of harmonic-space functions:
\beq
\bs{\mathcal{P}}[A(\bl_1), B(\bl_2)](\boldsymbol{\hat{n}}) \equiv \bs{\nabla}\mathcal{F}^{-1}[A(\bl_1)](\boldsymbol{\hat{n}}) \times \mathcal{F}^{-1}[B(\bl_2)](\boldsymbol{\hat{n}}).
\eeq
The single-pair estimators can all be written in the form 
\beq
\hat{\phi}_{XY}(\boldsymbol{\hat{n}}) = - \bs{\nabla} \cdot \mathcal{F}^{-1} \left[\lambda_{XY}(L) \mathcal{F}\left[\bs{\psi}_{XY}(\boldsymbol{\hat{n}}) \right]\right], \label{eq:phi-conf}
\eeq
with 
\barr
\bs{\psi}_{TT} &=& \bs{\mathcal{P}}\left[ \frac{\tilde{C}_{l_1}^{T \nabla T}}{C_{l_1}^{TT}} T(\bl_1), \frac{T(\bl_2)}{C_{l_2}^{TT}}\right],\\
\bs{\psi}_{EE} &=& \frac12 \sum_{\epsilon = \pm 1} \bs{\mathcal{P}}\left[\textrm{e}^{2 i \epsilon \varphi_{\bl_1}} \frac{\tilde{C}_{l_1}^{E \nabla E}}{C_{l_1}^{EE}}  E(\bl_1),  \textrm{e}^{-2 i \epsilon \varphi_{\bl_2}} \frac{E(\bl_2)}{C_{l_2}^{EE}} \right], ~~\\
\bs{\psi}_{TB} &=& \frac1{2i}  \sum_{\epsilon = \pm 1} \epsilon \bs{\mathcal{P}}\left[\textrm{e}^{2 i \epsilon \varphi_{\bl_1}} \frac{\tilde{C}_{l_1}^{T \nabla E}}{C_{l_1}^{TT}}  T(\bl_1),  \textrm{e}^{-2 i \epsilon \varphi_{\bl_2}} \frac{B(\bl_2)}{C_{l_2}^{BB}} \right],~~~\\
\bs{\psi}_{EB} &=& \frac1{2i}  \sum_{\epsilon = \pm 1} \epsilon \bs{\mathcal{P}}\left[\textrm{e}^{2 i \epsilon \varphi_{\bl_1}} \frac{\tilde{C}_{l_1}^{E \nabla E}}{C_{l_1}^{EE}}  E(\bl_1),  \textrm{e}^{-2 i \epsilon \varphi_{\bl_2}} \frac{B(\bl_2)}{C_{l_2}^{BB}} \right].~~~
\earr

These expressions are the flat-sky limit of the OH03 full-sky expressions.

\subsubsection{Case of $XY = TE$}
\label{sec:OH_B}

The separability property is not satisfied by the $TE$ estimator, due to the non-factorizable term in the denominator of $F_{TE}$ in Eq.~\eqref{eq:F_XY}. Instead of the optimal $F_{TE}$, one can use a slightly suboptimal coefficient, obtained by setting $C_l^{TE} = 0$ in Eq.~\eqref{eq:F_XY}, namely
\barr
F^{\rm eff}_{TE}(\bl_1, \bl_2) &=& \lambda^{\rm eff}_{TE}(L)\frac{f_{TE}(\bl_1, \bl_2)}{C_{l_1}^{TT} C_{l_2}^{EE}},  \\
\lambda^{\rm eff}_{TE}(L) &\equiv& \Bigg[\intL\frac{[f_{TE}(\bl_1, \bl_2)]^2}{C_{l_1}^{TT} C_{l_2}^{EE}} \Bigg]^{-1} \, .
\earr
The resulting suboptimal estimator $\hat{\phi}_{TE}^{\rm eff}$ also takes the form Eq.~\eqref{eq:phi-conf}, with
\barr
\bs{\psi}_{TE}^{\rm eff} &=& \frac12 \sum_{\epsilon = \pm 1} \bs{\mathcal{P}}\left[\textrm{e}^{2 i \epsilon \varphi_{\bl_1}} \frac{\tilde{C}_{l_1}^{T \nabla E}}{C_{l_1}^{TT}}  T(\bl_1),  \textrm{e}^{-2 i \epsilon \varphi_{\bl_2}} \frac{E(\bl_2)}{C_{l_2}^{EE}} \right] \nonumber\\
&&+\bs{\mathcal{P}}\left[\frac{\tilde{C}_{l_2}^{T \nabla E}}{C_{l_2}^{EE}} E(\bl_2), \frac{T(\bl_1)}{C_{l_1}^{TT}} \right].
\earr

\subsection{Optimal combination of single-pair estimators}

Given the five single-pair estimators $\hat{\phi}_\alpha({\bll})$ constructed for each $\alpha \in \{ TT, TE, EE, TB, EB \}$, HO02 combine them to form the estimator
\begin{equation}
\hat{\phi}_{\rm HO02}({\bll}) = \sum_{\alpha} w_\alpha(L) \hat{\phi}_\alpha({\bll})\,,
\label{eq:ho02_mv}
\end{equation}
where the optimal weights $w_\alpha(L)$ are obtained by minimizing the variance of the linear combination with the constraint that they sum up to unity i.e.~$\sum_\alpha w_\alpha = 1$. Subject to this constraint, one gets
\begin{equation}
w_\alpha = \frac{\sum_\beta ({\bs N}^{-1})_{\alpha\beta}} {\sum_{\beta\gamma} ({\bs N}^{-1})_{\beta\gamma}},
\label{eq:weights_ho02}
\end{equation}
where for each $L$, $\bs{N}_{\alpha \beta}(L)$ is the covariance matrix of the separate estimators $\hat{\phi}_{\alpha}$, whose elements are obtained by the generalization of Eq.~\eqref{eq:est-noise} to the cross-correlation of two estimators. The overall noise of this estimator is then $N_{\rm HO} \equiv \left(\sum_{\alpha\beta} ({\bs N}^{-1})_{\alpha\beta}\right)^{-1}$.

The final HO02 estimator thus takes the form 
\beq
\hat{\phi}_{\rm HO02}(\bll) = \intL  \sum_{XY} F_{XY}^{\rm HO02}(\bl_1, \bl_2) X(\bl_1) Y(\bl_2),\label{eq:HO02_final}
\eeq
where the sum runs over the five unique pairs $XY = TT, TE, EE, TB, EB$, and the weights are proportional to the single-pair optimal weights, each with a different proportionality coefficient:
\beq
F_{XY}^{\rm HO02}(\bl_1, \bl_2) = w_{XY}(L) F_{XY}(\bl_1, \bl_2). \label{eq:HO02-weights}
\eeq

The same procedure can be carried with the single-pair separable estimators of OH03. These estimators are all identical to the minimum-variance estimators of HO02, except for $\hat{\phi}_{TE}^{\rm eff}$, which is slightly suboptimal relative to $\hat{\phi}_{TE}$. Upon combining all five estimators, the OH03 estimator also takes the form of Eq.~\eqref{eq:HO02_final}, with weights
\beq
F_{XY}^{\rm OH03}(\bl_1, \bl_2) = w_{XY}^{\rm eff}(L) \lambda_{XY}^{\rm eff}(L) \frac{f_{XY}(\bl_1, \bl_2)}{(1 + \delta_{XY}) C_{l_1}^{XX} C_{l_2}^{YY}}. \label{eq:OH03-weights}
\eeq

\section{Global minimum-variance quadratic estimator} \label{sec:GMV}

It is easy to see that the final HO02 estimator is a linear combination of terms quadratic in $T, E, B$. Rather than splitting the optimization process in two steps, we instead directly seek the global minimum variance quadratic estimator, in one single step. By doing so, we can account for the correlations between different $XY$ pairs \emph{for each $(\bl_1, \bl_2)$}, rather than only after integrating over $(\bl_1, \bl_2)$, as done in the HO02 estimator. The estimator built this way is therefore necessarily less noisy than the HO02 estimator, as we will show explicitly.

\subsection{Harmonic-space expression}
\label{subsec:gmv}

We start by deriving the global-minimum-variance (hereafter GMV) estimator in harmonic space, following the steps of Appendix A of Hirata \& Seljak \cite{Hirata_03}. 


For each Fourier mode $\bl$, we define the three-dimensional vector $\bs{X}(\bl) = [T(\bl), E(\bl), B(\bl)]$. We seek an estimator of the form 
\beq
\hat{\phi}(\bll) = \intL X^i(\bl_1) \Xi_{ij}(\bl_1, \bl_2) X^j(\bl_2),
\label{eq:phi_gmv}
\eeq
where we use the Einstein summation convention. Without loss of generality, we may assume $\Xi_{ji}(\bl_2, \bl_1) = \Xi_{ij}(\bl_1, \bl_2)$, as only the part of the integrand symmetric under exchange of $(\bl_1, \bl_2)$ contributes to the integral.  

For $\bl_1 + \bl_2 \neq \bs{0}$, we define $f_{ij}(\bl_1, \bl_2)$ through 
\beq
\left\langle \frac{\delta}{\delta \phi(\bll)}\left(  X^i(\bl_1)X^j(\bl_2)\right)\right\rangle
  = \delta(\bl_1+\bl_2-\bll) f_{ij}(\bl_1,\bl_2).
\eeq
In other words, if $i = 1, 2, 3$ correspond to $X^i = T, E, B$, we have $f_{11}(\bl_1, \bl_2) = f_{TT}(\bl_1, \bl_2), f_{12}(\bl_1, \bl_2) = f_{TE}(\bl_1, \bl_2) = f_{21}(\bl_2, \bl_1)$, etc... Here again, we have $f_{ji}(\bl_2, \bl_1) = f_{ij}(\bl_1, \bl_2)$.

Requiring the estimator to be unbiased thus leads the constraint equation
\beq
\intL \Xi_{ij}(\bl_1, \bl_2) f_{ij}(\bl_1, \bl_2) = 1. \label{eq:constraint-hirata}
\eeq

Using the symmetry properties of $\Xi$, the variance of this estimator is then
\barr
N(\bll) &=& 2 \intL \Xi_{ij}(\bl_1, \bl_2) \Xi_{pq}(\bl_1, \bl_2) C_{l_1}^{ip} C_{l_2}^{jq}.
\label{eq:variance_gmv_n0}
\earr
Minimizing this variance under the constraint \eqref{eq:constraint-hirata} leads to 
\beq
C_{l_1}^{ip} \Xi_{pq} (\bl_1, \bl_2) C_{l_2}^{jq} = \frac{\lambda(L)}{2} f_{ij}(\bl_1, \bl_2), \label{eq:solution-hirata}
\eeq
where $\lambda(L)$ is a Lagrange multiplier. This equation is more easily solved in matrix form. For each $l$, we define the three by three symmetric matrix $[\bs{C}_l]$ with elements $C_l^{ij}$; similarly, for each pair $(\bl_1, \bl_2)$, we define the three by three matrices $[\bs{\Xi}(\bl_1, \bl_2)]$ and $[\bs{f}(\bl_1, \bl_2)]$. Equation \eqref{eq:solution-hirata} then has the solution
\beq
[\bs{\Xi}(\bl_1, \bl_2)] = \frac{\lambda(L)}{2} [\bs{C}_{l_1}]^{-1} [\bs{f}(\bl_1, \bl_2)] [\bs{C}_{l_2}]^{-1}.
\eeq
Inserting back into the constraint equation, we obtain
\beq
\lambda(L)^{-1} = \intL \frac12 \textrm{Tr}\left([\bs{C}_{l_1}]^{-1} [\bs{f}(\bl_1, \bl_2)] [\bs{C}_{l_2}]^{-1} [\bs{f}(\bl_2, \bl_1)] \right).
\eeq
The noise of the minimum-variance estimator is then simply $N(L) = N_{\rm GMV} (L) = \lambda(L)$.

Putting everything together, the GMV estimator takes the final form 
\barr
\hat{\phi}_{\rm GMV}(\bll) &=&  \intL  \sum_{XY} F_{XY}^{\rm GMV}(\bl_1, \bl_2) X(\bl_1) Y(\bl_2),
\label{eq:gmv_est}
\earr
where the sum runs over the five unique pairs $XY = TT, TE, EE, TB, EB$. Explicitly, the weights are 
\barr
&&F_{TT}^{\rm GMV}(\bl_1, \bl_2)  = \frac{N_{\rm GMV}(L)}{2 D_{l_1} D_{l_2}} \times  \nonumber\\
&&~~~ \Big{[}C_{l_1}^{EE} C_{l_2}^{EE} f_{TT}(\bl_1, \bl_2) + C_{l_1}^{TE} C_{l_2}^{TE} f_{EE}(\bl_1, \bl_2) \nonumber\\
 && ~~~~~ -  C_{l_1}^{EE} C_{l_2}^{TE} f_{TE}(\bl_1, \bl_2) 
 -   C_{l_2}^{EE} C_{l_1}^{TE} f_{TE}(\bl_2, \bl_1)\Big{]}, ~~~~~\\
&& F_{EE}^{\rm GMV}(\bl_1, \bl_2) =\frac{N_{\rm GMV}(L)}{2 D_{l_1} D_{l_2}} \times \nonumber\\
&&~~~\Big{[}   C_{l_1}^{TE} C_{l_2}^{TE} f_{TT}(\bl_1, \bl_2) 
+  C_{l_1}^{TT} C_{l_2}^{TT} f_{EE}(\bl_1, \bl_2) \nonumber\\
 &&~~~~~-   C_{l_1}^{TE} C_{l_2}^{TT} f_{TE}(\bl_1, \bl_2)-  C_{l_2}^{TE} C_{l_1}^{TT} f_{TE}(\bl_2, \bl_1) \Big{]},\\
&& F_{TE}^{\rm GMV}(\bl_1, \bl_2) = \frac{N_{\rm GMV}(L)}{D_{l_1} D_{l_2}} \times \nonumber\\
&& ~~~\Big{[} - C_{l_1}^{TE} C_{l_2}^{EE} f_{TT}(\bl_1, \bl_2) 
- C_{l_1}^{TE} C_{l_2}^{TT} f_{EE}(\bl_1, \bl_2) \nonumber\\
 &&~~~~~+ C_{l_1}^{EE} C_{l_2}^{TT} f_{TE}(\bl_1, \bl_2) 
+ C_{l_1}^{TE} C_{l_2}^{TE} f_{TE}(\bl_2, \bl_1) \Big{]},\\
&& F_{TB}^{\rm GMV}(\bl_1, \bl_2) = \frac{N_{\rm GMV}(L)}{D_{l_1} C_{l_2}^{BB}} \times \nonumber\\
&& ~~~\Big{[} C_{l_1}^{EE} f_{TB}(\bl_1, \bl_2) - C_{l_1}^{TE} f_{EB}(\bl_1, \bl_2)\Big{]},\\
&& F_{EB}^{\rm GMV}(\bl_1, \bl_2) = \frac{N_{\rm GMV}(L)}{D_{l_1} C_{l_2}^{BB}} \times \nonumber\\
&&~~~\Big{[} - C_{l_1}^{TE} f_{TB}(\bl_1, \bl_2) + C_{l_1}^{TT} f_{EB}(\bl_1, \bl_2)\Big{]} \, ,
\earr
where $D_l \equiv C^{TT}_l C^{EE}_l - [C^{TE}_l]^2$. 

These explicit expressions should make it very clear that the GMV estimator is different from the HO02 estimator. Put differently, the first step of the likelihood-based iterative technique of Ref.~\cite{Hirata_03} is \emph{not} equivalent to the HO02 estimator. Indeed, in the HO02 estimator, the weight $F_{XY}$ of each pair $XY$ is proportional to $f_{XY}$ only (times a function of $L$), even after combining all single-pair estimators; in contrast, for the GMV estimator, the weight of each pair is a linear combination of the response coefficients from \emph{all} pairs. The weights in the GMV estimator would not separately minimize the variance of an individual $XY$ pair, but they provide a global optimum when combining all the pairs together.

These expressions also show that the weights are all sums of products of function of $\bl_1$ with functions of $\bl_2$, including $F_{TE}^{\rm GMV}$. In other words, the GMV estimator is separable without requiring any additional approximation, making it well adapted for efficient computations, as we now discuss.

As a side note, let us point out that the GMV estimator (just like the HO02 an OH03 estimators) can be split into two pieces, built from $\{TT, TE, EE\}$ and $\{TB, EB\}$, respectively, which are uncorrelated as $C^{TB}_\ell = C^{EB}_\ell = 0$. Our publicly available Python code \texttt{GlobalLensQuest} first computes these two separate uncorrelated estimators $\hat{\phi}_a \equiv \hat{\phi}_{\{TT, TE, EE\}}$ and $\hat{\phi}_b \equiv \hat{\phi}_{\{TB, EB\}}$, and then combines them with inverse variance weighting to obtain the GMV estimator.


\subsection{Compact configuration-space expression}

We may write the GMV estimator in even more compact form by defining the inverse-covariance-weighted fields
\beq
\overline{\bs{X}}(\bl) \equiv [\bs{C}_{l}]^{-1} \bs{X}(\bl) ,
\label{eq:inv-var-weight}
\eeq
and write 
\beq
\overline{T}(\bl_1)\equiv \overline{X}^1(\bl_1) = [C_{l_1}^{EE} T(\bl_1) - C_{l_1}^{TE} E(\bl_1)]/D_{l_1},  
\eeq
and similarly $\overline{E}(\bl_1) = \overline{X}^2, \overline{B}(\bl_1) = \overline{X}^3$. We then have
\beq
\hat{\phi}_{\rm GMV}(L) = \frac{\lambda(L)}2 \intL \overline{X}^i(\bl) f_{ij}(\bl_1, \bl_2) \overline{X}^j(\bl_2).
\eeq
We moreover define the Wiener-filtered fields 
\beq
X^i_{\rm WF}(\bl) \equiv \widetilde{C}_l^{ij} ~\overline{X}^j,
\eeq
where $\widetilde{C}_l^{11} \equiv \widetilde{C}_l^{T \nabla T}, \widetilde{C}_l^{12} = \widetilde{C}_l^{21}  \equiv \widetilde{C}_l^{T \nabla E}$, etc., and write 
\beq
T_{\rm WF}(\bl) \equiv X^1_{\rm WF}(\bl) \equiv \widetilde{C}_l^{T \nabla T~} \overline{T}(\bl) + \widetilde{C}_l^{T \nabla E}~ \overline{E}(\bl),
\eeq
and similarly for $E_{\rm WF}(\bl) \equiv X^2_{\rm WF}(\bl)$.

In terms of these fields, the GMV estimator takes the particularly simple form
\barr
\hat{\phi}_{\rm GMV}(\bll) = \frac{\lambda(L)}{2} \intL (\bll \cdot \bl_1)\Big{[} 2 T_{\rm WF}(\bl_1) \overline{T}(\bl_2)  \nonumber\\
+  \sum_{\epsilon = \pm 1}  E_{\rm WF}(\bl_1) \textrm{e}^{- 2 i \epsilon \varphi_{\bl_1}} \left(\overline{E}(\bl_2) + i \epsilon \overline{B}(\bl_2) \right) \textrm{e}^{2 i \epsilon \varphi_{\bl_2}} \Big{]}.~
\earr
This form is well adapted for efficient evaluation, as it is the sum of  convolutions of functions of $\bl_1$ with functions of $\bl_2$. To see this, let us define
\barr
~_{\pm 2} E_{\rm WF}(\bl) &\equiv& E_{\rm WF}(\bl) \textrm{e}^{\pm 2 i  \varphi_{\bl}}, \\
~_{\pm 2} \overline{P}(\bl) &\equiv&  {\frac 12}\left(\overline{E}(\bl) \pm  i \overline{B}(\bl) \right) \textrm{e}^{\pm 2 i \varphi_{\bl}}.
\earr
We may then express the GMV estimator in terms of the configuration-space versions of these fields (i.e.~their inverse-Fourier transforms):
\beq \label{eq:nabpsi}
\hat{\phi}_{\rm GMV}(\boldsymbol{\hat{n}}) = - \bs{\nabla} \cdot \mathcal{F}^{-1}\left[\lambda(L) \mathcal{F}[ \bs{\psi}_{\rm GMV}(\boldsymbol{\hat{n}})] \right], \\
\eeq
where
\barr \label{eq:psi}
\bs{\psi}_{\rm GMV}(\boldsymbol{\hat{n}}) &=& \bs{\nabla} T_{\rm WF}(\boldsymbol{\hat{n}})~\overline{T}(\boldsymbol{\hat{n}})  \nonumber\\
&& + \sum_{s = \pm 2} \bs{\nabla}[_{-s} E_{\rm WF}(\boldsymbol{\hat{n}})] ~_{s} \overline{P}(\boldsymbol{\hat{n}}).
\earr
This expression is the flat-sky equivalent of Eq.~(3) in Ref.~\cite{Planck_lensing_18}, derived in Ref.~\cite{Carron_19}. 
A similar expression is derived in Ref.~\cite{Peloton_17}, in terms of $(T, Q, U)$ rather than $(T, E, B)$; nevertheless, it is also incorrectly stated in that paper that this estimator is identical to an estimator built out of single-pair estimators, i.e.~the HO02 estimator. 

\section{Suboptimal quadratic estimator (SQE) used in recent data analyses} \label{sec:SQE}

While the full expression for the configuration-space GMV estimator was already known (although it was not known that it differs from the HO02 estimator) \cite{Planck_lensing_18, Carron_19}, in practice only an approximate version was used for data analyses thus far. Instead of using the full covariance matrix $[\bs{C}_l]$ in Eq.~\eqref{eq:inv-var-weight}, the Planck collaboration \cite{Planck_lensing_16, Planck_lensing_18} and SPT collaboration \cite{Wu_19}  
approximate it as diagonal by setting $C_l^{TE} = 0$ -- note that \cite{Planck_lensing_18} still use the exact response coefficients $f_{XY}(\bl_1, \bl_2)$. This simplification allows to deal with a cut-sky setup with a lower computational cost; it moreover preserves the configuration space separability. We denote this suboptimal quadratic estimator SQE. Explicitly, the weights of this estimator are
\barr
F_{XY}^{\rm SQE}(\bl_1, \bl_2) &=& \lambda_{\rm SQE}(L) \frac{f_{XY}(\bl_1, \bl_2)}{(1 + \delta_{XY}) C_{l_1}^{XX} C_{l_2}^{YY}} , \label{eq:weight-SQE}\\
\lambda_{\rm SQE}(L) &\equiv& \left( \intL \sum_{XY} \frac{f_{XY}(\bl_1, \bl_2)^2}{(1 + \delta_{XY}) C_{l_1}^{XX} C_{l_2}^{YY}} \right)^{-1},~~
\earr
where again the sum runs of the five distinct pairs $XY$. 

By definition, this estimator is suboptimal relative to the GMV estimator. Furthermore, it should be clear from Eq.~\eqref{eq:OH03-weights} that it is also noisier than the OH03 estimator (and as we will see, also noisier than the HO02 estimator). Indeed, the OH03 estimator accounts for the covariances between different single-pair estimators, which depend on $C_l^{TE}$; as a consequence, the $L$-dependent proportionality constant in Eq.~\eqref{eq:OH03-weights} is different for each pair. In contrast, the SQE amounts to neglecting correlations between single-pair estimators, and simply using their inverse-variance combination, leading to the same coefficient $\lambda_{\rm SQE}(L)$ for all weights in Eq.~\eqref{eq:weight-SQE}. Given that the SQE estimator is effectively a linear combination of single-pair estimator, and that the OH03 weights represent the optimal linear combination of single-pair estimators, we conclude that the SQE estimator must be suboptimal to the OH03 estimator.

To compute the noise of this estimator, we must account for $C_l^{TE} \neq 0$. We find
\barr
N_{\rm SQE}(L) = \lambda_{\rm SQE}^2(L) \intL \sum_{XY} \frac{f_{XY}(\bl_1, \bl_2)}{(1 + \delta_{XY}) C_{l_1}^{XX} C_{l_2}^{YY}} \nonumber\\
\sum_{UV} \left[ \frac{f_{UV}(\bl_1, \bl_2) C_{l_1}^{XU} C_{l_2}^{YV}}{(1 + \delta_{UV}) C_{l_1}^{UU} C_{l_2}^{VV}} + \frac{f_{UV}(\bl_2, \bl_1) C_{l_1}^{XV} C_{l_2}^{YU}}{(1 + \delta_{UV}) C_{l_2}^{UU} C_{l_1}^{VV}}\right].~~
\earr
The SQE estimator enables faster evaluation with a cut-sky, at the cost of only $\sim 3\%$ increase in the reconstruction noise for \textit{Planck} \citep{Planck_lensing_18}, as we confirm in Fig.~\ref{fig:mult_ratio_tmv_HO02}. However, we will see that for more sensitive experimental setups, the reconstruction noise penalty can be more than $10\%$ and thus a full joint filtering analysis of the temperature and polarization maps would be beneficial in future.

\section{Quantitative comparison of different quadratic estimators} \label{sec:comparison}

\subsection{Experimental setups} \label{sec:setups}

In order to evaluate the variance of different quadratic estimators, we use three different setups which correspond to the \textit{Planck}, SO-like, and CMBS4-like experiments. In Table~\ref{tab:exp}, we provide the adopted specifications for these setups. The Gaussian random noise of the detector is calculated as 
\begin{equation}
    C^{T, E/B}_\ell|_\mathrm{noise} = (\Delta_{T, P})^2 e^{\ell(\ell + 1) \sigma^2 /8 \ln{2}}
\end{equation}
where $\Delta_{T,P}$ denote the white-noise of the detector in $\mu$K-radian, and $\sigma$ is the full width at half maximum (FWHM) of the beam in arcmin.

\begin{table}[h]
\begin{tabular}{|c|c|c|c|c|c}
\hline
Experiment & $\ell_\mathrm{max}$ & $\Delta_T$ & $\Delta_P$ & $\sigma$ \vsp
 & & $\mu\mathrm{K}$-arcmin & $\mu\mathrm{K}$-arcmin & arcmin \vsp
\hline
\textit{Planck} & 3000 & 35.0 & 60.0 & 5.0  \vsp
SO & 3000 & 8.0 & 8.0$\sqrt{2}$ & 1.4  \vsp
CMBS4 & 3000 &  1.0 & 1.0$\sqrt{2}$ & 1.0  \vsp
\hline
\end{tabular}
\centering \caption{Experimental specifications used in this work.}
\label{tab:exp}
\end{table}

It has been shown that  extra-galactic foregrounds can bias the CMB lensing reconstruction from temperature maps and different strategies have been proposed to mitigate this issue \citep[e.g.][]{Ferraro_18, Schaan_19, Madhavacheril_18, Darwish_21, Engelen_14}. We neglect the foregrounds in this study, and simply choose $\ell_\mathrm{max} = 3000$ in both temperature and polarization. Although it is possible to go for a much higher $\ell_\mathrm{max}$ in polarization than in temperature due to lack of strongly polarized foregrounds, for simplicity we take $\ell^T_\mathrm{max} = \ell^P_\mathrm{max}$. We have checked our results with different $\ell_\mathrm{max}$ ranges and found no drastic difference in our results.

\subsection{Comparison of reconstruction noises}
\label{subsec:comparison_disc}

\begin{figure}[ht]
\centering
\includegraphics[width=9cm]{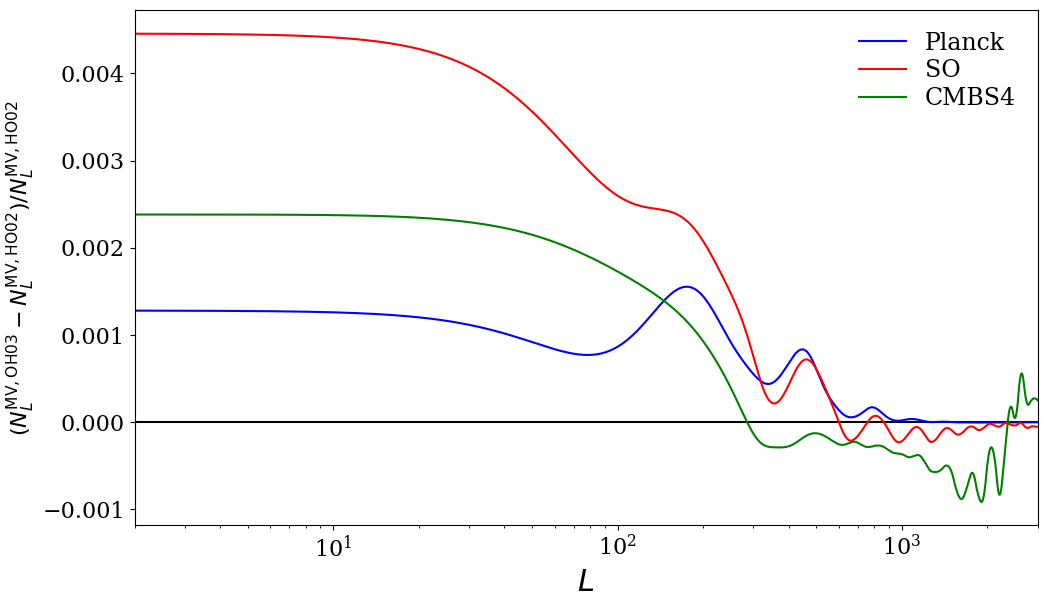}
\centering \caption{Fractional difference between the variance of the OH03 and HO02 estimators. Different colors correspond to the different experimental setups described in Sec.~\ref{sec:setups}.}
\label{fig:mult_OH03_HO02}
\end{figure}

We start by comparing the OH03 and HO02 quadratic estimators in Figure~\ref{fig:mult_OH03_HO02}. We find that the HO02 estimator is systematically less noisy than the OH03 estimator for large angular scales (by less than 0.5\%). Interestingly, the OH03 estimator becomes slightly less noisy than the HO02 estimator for $L$ larger than several hundreds. We have checked that our numerical integrals are converged to better than 0.01\% relative accuracy up to $L \approx 2000$, and to better than 0.03\% for $L \lesssim 3000$ which we also show in Fig.~\ref{fig:convergence}. This gives us confidence that the $\sim 0.08\%$ improvement of OH03 estimator over HO02 seen for a CMBS4-like experiment is real and not a numerical artifact.

The lower noise of OH03 at small angular scales may appear surprising at first, given that this estimator uses a TE estimator suboptimal to that of HO02. However, the suboptimal TE estimator does not guarantee that the overall OH03 estimator (obtained by optimally combining the five single-pair estimators) is noisier than the overall HO02 estimator: indeed, if the OH03 TE estimator happens to be more correlated with the TT and EE estimators than the HO02 TE estimator which is the case here, the overall combination of OH03 estimators can be less noisy than that of HO02.

\begin{figure}[ht]
\centering
\includegraphics[width=9cm]{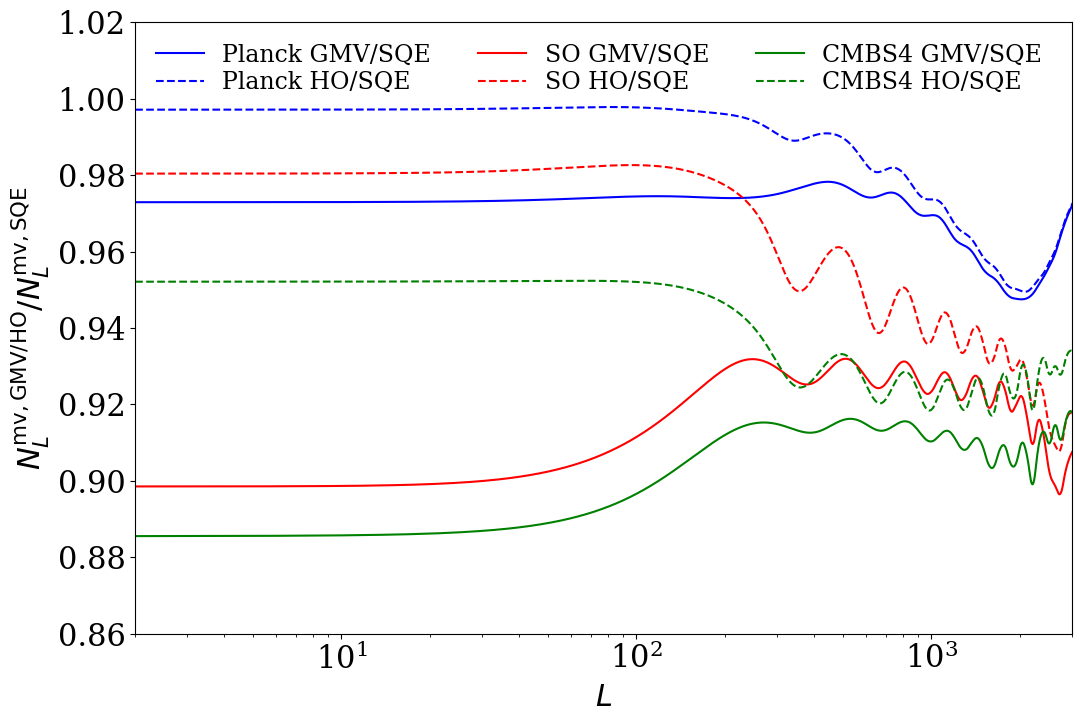}
\centering \caption{Ratio of the minimum variance reconstruction noise of the GMV and SQE estimators, and HO02 and SQE estimators for different experimental setups.}
\label{fig:mult_ratio_tmv_HO02}
\end{figure}

In Fig.~\ref{fig:mult_ratio_tmv_HO02}, we show the ratios of the reconstruction noise of the GMV and HO02 estimators to that of the SQE estimator for different experimental setups. As expected, we find that the variance of the GMV quadratic estimator is lower than that of the HO02 and SQE estimators. Also, the variance of HO02 estimator is smaller than the SQE estimator. For a \textit{Planck}-like experimental setup, on large angular scales ($L \lesssim 500$), the difference between the SQE and GMV estimators is of the order of 3\% and can reach $\sim 6\%$ around $L \sim 2000$. However, for more sensitive experiments, this difference reaches $\sim 11\%$ and $\sim 12\%$ at $L \lesssim 100$ and $L \sim 2000$ for SO- or CMBS4-like experiments, respectively. This result may motivate using the full covariance matrix $[\bs{C}_l]$ in Eq.~\eqref{eq:inv-var-weight} instead of assuming $C_l^{TE} = 0$, in order to obtain more precise results in future data analyses.  


From Fig.~\ref{fig:mult_ratio_tmv_HO02}, we can also see that for a \textit{Planck}-like experimental setup, at small angular scales ($L \gtrsim 1000$) the GMV and HO02 estimators perform almost equally well, while for SO- and CMBS4-like experiment GMV  outperforms HO02 everywhere. On large angular scales, the difference between the GMV and HO02 is much more significant for all the experiments considered here. 
For \textit{Planck}, polarization noise is significant, so the difference in lensing estimators is quite small on all scales. However, although CMBS4-like experiments are $EB$-dominated for the purpose of lensing reconstruction, the improvement of the GMV over HO02 is driven by a significant difference in the TT-TE-EE part of the minimum-variance estimator (rather than improved filtering of $E$ improving $EB$; see Appendix~\ref{app:gmv_ho02}). The effect on the combined MV estimator is largest at $L \lesssim 200$ where the MV estimator has the largest contribution from TT-TE-EE.

\section{Conclusions} \label{sec:conclusion}

Quadratic estimators (QEs) are widely used to reconstruct the CMB lensing potential from CMB temperature and polarization maps. In fact, up until the very recent POLARBEAR \cite{Adachi_20} and SPTPol \cite{Millea_20b} results, all maps of the CMB lensing potential have been constructed using QEs. In this work, we present a clear comparison between different QEs, both in terms of explicit equations, and quantitatively, by comparing their reconstruction noise. Importantly, we show that the Hu-Okamoto \cite{Hu_02} (HO02) optimization method, consisting in first constructing optimal single-pair quadratic estimators, and then their optimal linear combination, does not lead to the absolute minimum-variance QE. Instead, we derive the global-minimum-variance (GMV) QE, which minimizes the variance of quadratic temperature and polarization combinations in one single step. 



Interestingly, the GMV estimator derived here had been hiding in plain sight in previous works. It is equivalent to the first step (or weak-signal limit) of likelihood-based methods \cite{Hirata_03, Peloton_17, Carron_17}, which is therefore \emph{not} equivalent to the HO02 estimator, contrary to what was previously thought (although technically the GMV estimator with non-perturbative lensed gradient weights presented here is a modification to the first-step of likelihood based estimators so that the result is non-perturbatively unbiased). Our work is the first to note that the HO02 estimator is not the global-optimum quadratic estimator, and make this point sharply clear through explicit expressions, as well as numerical comparisons. Indeed, we show that the reconstruction noise of the GMV estimator is lower than that of the HO02 estimator by up to $\sim 9\%$ for a SO-like experiment. 

We also study the suboptimal QE used in the 2018 Planck \cite{Planck_lensing_16, Carron_17, Planck_lensing_18} and recent SPT \cite{Wu_19} lensing analyses (SQE), which is obtained from the GMV quadratic estimator (appropriately generalized to account for beam and pixel convolution), with the additional approximation of neglecting $C_{l}^{TE} = 0$ in the inverse filter matrix. We show that this approximation makes the SQE suboptimal not only relative to the GMV estimator, but also relative to the HO02 and OH03 estimators. We evaluate the reconstruction noise of the different estimators for ongoing and planned CMB experimental setups and find that while the improvement in the reconstruction noise between the SQE and HO02 estimator is of order $\sim 1-8\%$, the difference between the SQE and GMV estimators is $\sim 9-12\%$ for more sensitive experiments, especially on large angular scales $L \lesssim 10^2$ and scales around $L \sim 2000$. This improvement amounts to achieving a better sensitivity for the same experiment at no additional cost. This should motivate overcoming the added complexity associated with joint filtering of cut-sky temperature and polarization maps, in order to be able to use the GMV estimator in future lensing data analyses. Our Python code \texttt{GlobalLensQuest}\footnote{\url{https://github.com/abhimaniyar/GlobalLensQuest}} to compare and compute the noise variances of the HO02, OH03, GMV, and SQE estimators and Julien Carron's codes \texttt{LensIt}\footnote{\url{https://github.com/carronj/LensIt}} and \texttt{plancklens}\footnote{\url{https://github.com/carronj/plancklens}} which can perform the optimal GMV operation with anisotropic noise and cut-sky are publicly available.


While in this work we have chosen to present relevant equations in the flat-sky limit for conciseness, it is straightforward to generalize our results to the full-sky case. The flat-sky fields $X(\bl)$ are to be replaced by the full-sky harmonic coefficients $X_{\ell m}$; the generalization of Eq.~\eqref{eq:cmbavg} and of the coupling coefficients $f_{XY}(\bl, \bl')$ are then provided in Ref.~\cite{Hu_03}. While \cite{Hu_03} provide the full-sky expressions for the HO02 estimator, \cite{Carron_19} provide the full-sky expressions for the GMV estimator incorporating the instrumental beam response and anisotropic noise. We expect a comparable improvement over the full-sky version of the HO02 estimator \cite{Hu_03} when using the GMV estimator \cite{Carron_19}. 

The approach presented here for the GMV estimator would also apply to any other joint estimator constructed from similar linear combinations of other estimators. For example, the foreground-immune hybrid QE of Ref.~\cite{Schaan_19} splits the $TT$ lensing estimator into magnification-only and shear-only estimators, and then form a hybrid estimator through a minimum-variance linear combination of these two estimators. Their hybrid estimator can be further optimized by following the logic presented here, i.e.~searching for the global-minimum-variance shear and magnification estimator, accounting for correlations for each $(\bl_1, \bl_2)$, rather than after integration over $\bl_1, \bl_2$. 

For SO- and CMBS4-like experiments, on large angular scales ($L < 100$) the reconstruction is expected to be signal dominated \cite{Ade_19, S4_19}. 
The power spectrum $C^{\phi \phi}_L$ uncertainty is therefore dominated by cosmic variance and using the GMV estimator rather than the HO02 or SQE estimators would
not drastically affect the measurement of $C^{\phi \phi}_L$ on these large angular scales. However, the reduction in the noise of the reconstructed $\phi$ field on the signal-dominated large angular scales will be beneficial for science goals which involve cross-correlation of the $\phi$ field with other tracers of large-scale structure \cite{Ade_19, Schmittfull_18}, utilizing the sample variance cancellation through cross-correlations. Also, lensing induced B-modes act as a source of noise and limit the measurement of the primordial B-modes \cite{Zaldarriaga_03} which is a major scientific goal for CMB experiments. These modes can be removed using map-level estimates of both the primordial E-modes and lensing potential $\phi$ with a technique called delensing and depend on the estimate of the particular realization of $\phi$ in the given patch of the sky \cite{Seljak_04, Ade_19, S4_19}. Lower noise estimates of the $\phi$ field will therefore be crucial for such operations and motivate using the GMV estimator instead of other QEs.

Even if future CMBS4 experiments will likely use likelihood-based iterative methods to reconstruct the lensing potential, QEs will remain very useful as a forecasting and cross-checking tool. More immediately, QEs are still the primary lensing reconstruction tool for current and near-future CMB experiments. The GMV estimator will therefore be a useful tool to harvest even more information out of the CMB data.

\section*{Acknowledgements}

We acknowledge useful discussions with Colin Hill, Emmanuel Schaan, and Simone Ferraro. YAH is supported by NSF award No 1820861 and NASA grant No 80NSSC20K0532. JC acknowledges support from a SNSF Eccellenza Professorial Fellowship (No. 186879). AL has support from the UK STFC grant ST/T000473/1.

\onecolumngrid
\appendix
\section{Explicit form for the GMV $N^{(1)}(L)$}
\label{app:NOne}
As pointed out in Sec.~\ref{sec:notation}, in the main text we only optimize relative to the reconstruction noise $N^{(0)}(L)$. Here we give an explicit expression for the additional $N^{(1)}(L)$ bias~\cite{Kesden_03} for the GMV estimator, for which an explicit expression has not been provided in the literature.

The covariance of the GMV estimator given by Eq.~\eqref{eq:phi_gmv} becomes
\barr
\langle \hat{\phi}(\bll) \hat{\phi}(\bll') \rangle = \intL \intLp && \langle X^i(\bl_1) X^j(\bl_2) X^p(\bl'_1) X^q(\bl'_2) \rangle  \Xi_{ij}(\bl_1, \bl_2) \Xi_{pq}(\bl'_1, \bl'_2) \, .
\label{eq:var_gmv_general}
\earr
Evaluating the expectation value in the brackets
\barr
\langle X^i(\bl_1) X^j(\bl_2) X^p(\bl'_1) X^q(\bl'_2) \rangle &=& (2\pi)^4 \big[C^{ij}_{l_1}C^{pq}_{l'_1} \delta_{\rm D}(\bll)\delta_{\rm D}(\bll') + C^{ip}_{l_1}C^{jq}_{l'_1}\delta_{\rm D}(\bl_1+\bl'_1)\delta_{\rm D}(\bl_2+\bl'_2) \nonumber \\
&&+ C^{iq}_{l_1}C^{jp}_{l'_1}\delta_{\rm D}(\bl_1+\bl'_2)\delta_{\rm D}(\bl_2+\bl'_1)    \big] + (2\pi)^2 T^{ijpq}(\bl_1, \bl_2, \bl'_1, \bl'_2)\delta_{\rm D}(\bll + \bll') \, ,
\label{eq:vargmv_n1}
\earr
where the first term in the square brackets disappears because $\bll \neq 0$ and rest of the terms in the square bracket represent $N^{(0)}(L)$ and $T^{ijpq}(\bl_1, \bl_2, \bl'_1, \bl'_2)$,  the trispectrum containing terms that contribute to the lensing power spectrum signal and the signal-dependent $N^{(1)}(L)$ bias. Following \cite{Kesden_03}, the trispectrum term can be written in terms of $f_{ij}(\bl_1, \bl_2)$ 
to first order in explicit $C^{\phi \phi}_L$ as
\barr
T^{ijpq}(\bl_1, \bl_2, \bl'_1, \bl'_2) = C^{\phi \phi}_{|\bl_1+\bl_2|}f_{ij}(\bl_1, \bl_2) f_{pq}(\bl'_1, \bl'_2) + C^{\phi \phi}_{|\bl_1+\bl'_1|}f_{ip}(\bl_1, \bl'_1) f_{jq}(\bl_2, \bl'_2) + C^{\phi \phi}_{|\bl_1+\bl'_2|}f_{iq}(\bl_1, \bl'_2) f_{jp}(\bl_2, \bl'_1).~~~~
\label{eq:gmvtrispectrum}
\earr

Substituting Eqs.~\eqref{eq:vargmv_n1} and \eqref{eq:gmvtrispectrum} in Eq.~\eqref{eq:var_gmv_general} we have
\beq
\langle \hat{\phi}(\bll) \hat{\phi}(\bll') \rangle = (2\pi)^2 \delta_{\rm D}(\bll+\bll') [C^{\phi \phi}_L + N^{(0)}(L) + N^{(1)}(L)]\,,
\label{eq:est-noise_n1}
\eeq
where 
\barr
N^{(1)}(L) &\equiv& \intL \intLp \Xi_{ij}(\bl_1, \bl_2) \Xi_{pq}(\bl'_1, \bl'_2) \times \big[ C^{\phi \phi}_{|\bl_1+\bl'_1|}f_{ip}(\bl_1, \bl'_1) f_{jq}(\bl_2, \bl'_2) + C^{\phi \phi}_{|\bl_1+\bl'_2|}f_{iq}(\bl_1, \bl'_2) f_{jp}(\bl_2, \bl'_1) \big] \, ,
\earr

The optimal weight matrix $\Xi_{ij}(\bl_1, \bl_2)$ was determined in order to minimize the variance while only considering $N^{(0)}(L)$.

\begin{figure}
\centering
\begin{subfigure}{.5\textwidth}
  \centering
  \includegraphics[width=\textwidth]{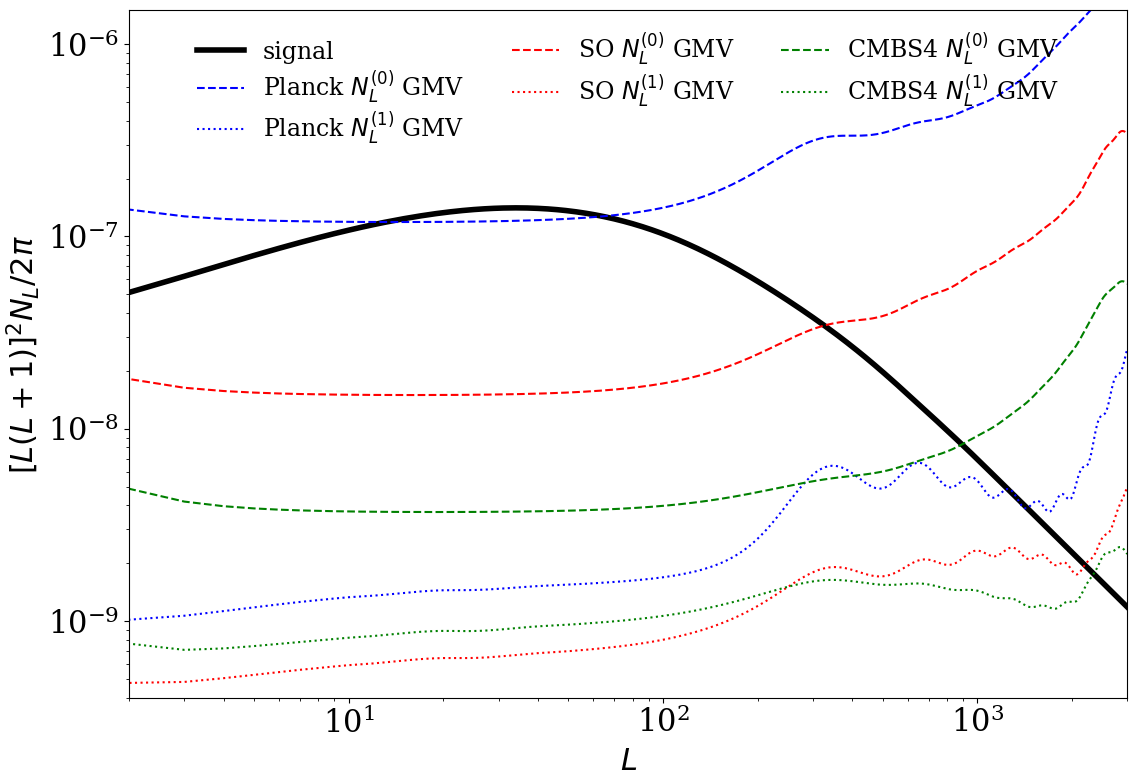}
\end{subfigure}%
\begin{subfigure}{.5\textwidth}
  \centering
  \includegraphics[width=\textwidth]{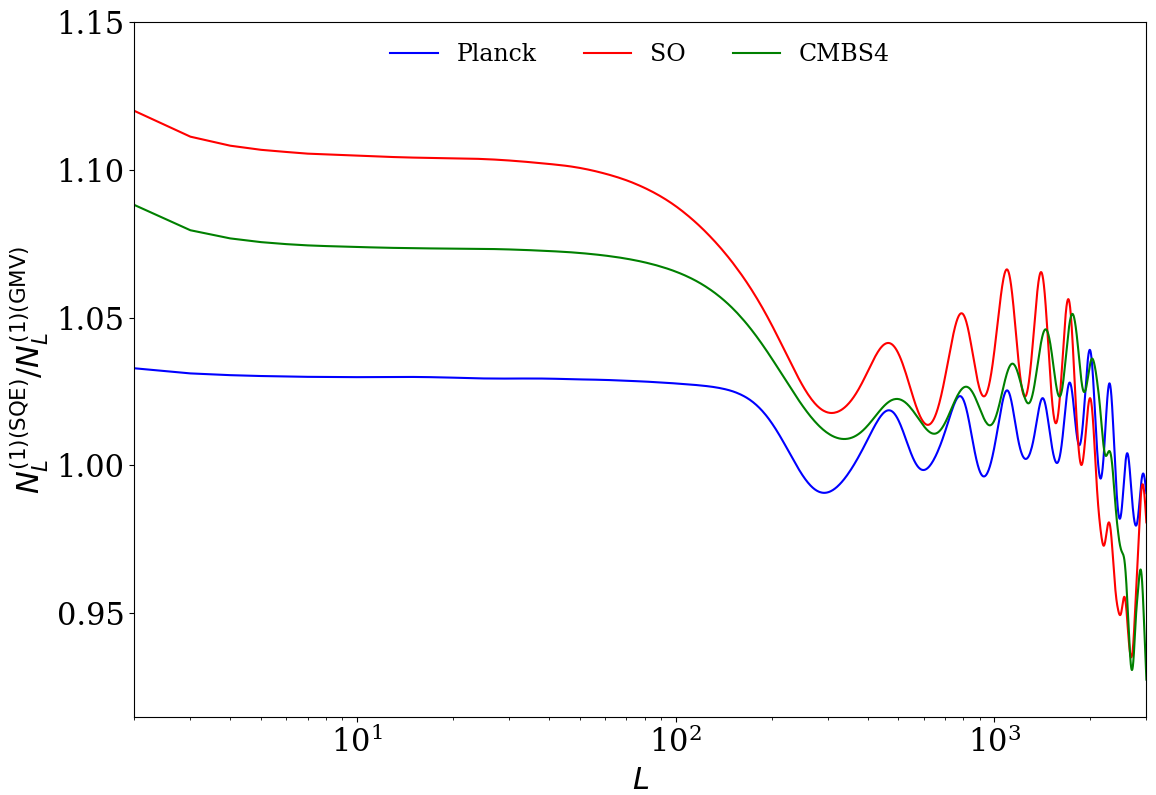}
\end{subfigure}
\caption{Comparison of the $N^{(0)}(L)$ and $N^{(1)}(L)$ curves for the GMV estimator (left) and ratio of the $N^{(1)}(L)$ for SQE and GMV estimators (right) for given experimental configurations.
}
\label{fig:mult_gmv_n0n1_sqe}
\end{figure}

In the left panel of Fig.~\ref{fig:mult_gmv_n0n1_sqe}, we show the comparison between the $N^{(0)}(L)$ and $N^{(1)}(L)$ curves for the GMV estimator for our given experimental configurations. We find that for \textit{Planck}-like experiments $N^{(1)}(L)$ is a couple of orders of magnitude smaller than $N^{(0)}(L)$, while for more sensitive (less noisy) SO- and CMBS4-like experiments, $N^{(1)}(L)$ is a factor of few to an order of magnitude smaller than $N^{(0)}(L)$. On small scales it can however become comparable to the signal spectrum and is important to model in any likelihood analysis. It would be straightforward to apply the perturbative likelihood approximation of \cite{Planck_lensing_16,Planck_lensing_18} (which accounts consistently for the signal dependence of $N^{(1)}(L)$) to the GMV estimator. In the right panel of Fig.~\ref{fig:mult_gmv_n0n1_sqe}, we show the ratio of the $N^{(1)}(L)$ for SQE and GMV estimators for different experiments considered here. For SO- and CMBS4-like experiments, $N^{(1)}(L)$ bias for the GMV estimator is smaller than for the SQE estimator for $L \lesssim 1800$. For \textit{Planck}-like experiments, apart from the large angular scales $L \lesssim 200$ where GMV estimator gives smaller $N^{(1)}(L)$ bias than the SQE estimator, both the estimators have almost the same $N^{(1)}(L)$ bias.

\section{Numerical convergence}
\begin{figure}
\centering
\includegraphics[width=\textwidth]{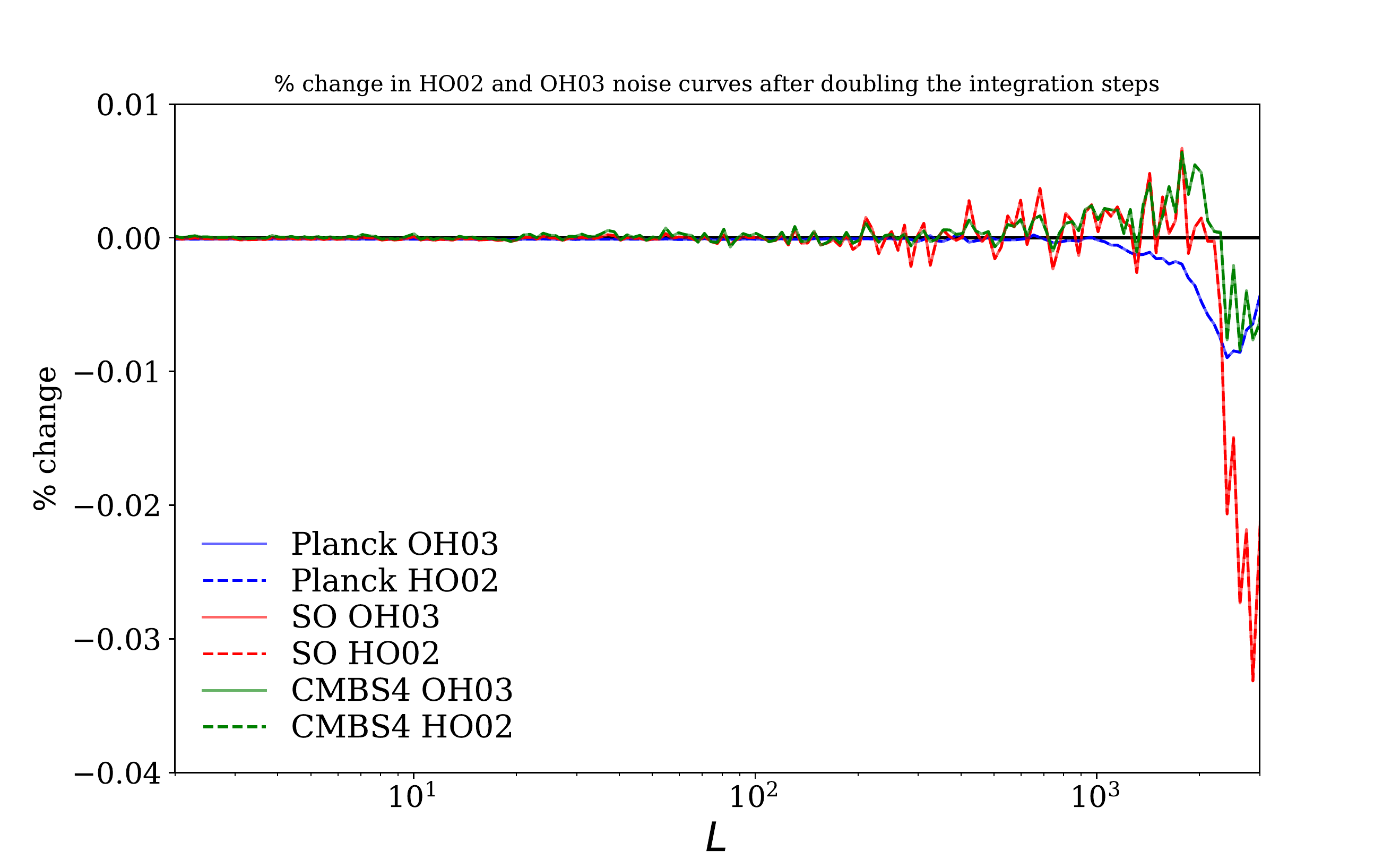}
\centering \caption{\% change in the noise curves for HO02 and OH03 estimators when we double the number of integration steps in our Python code.}
\label{fig:convergence}
\end{figure}

In Fig.~\ref{fig:convergence}, we show the result of the convergence test we perform for our Python code. It shows the \% change in the noise calculation of HO02 (dashed curves) and OH03 (solid curves) estimators when we double the number of steps in the angular part of the integration for the given $\ell_{\rm max}$ and $L \lesssim 3000$. The dashed HO02 curves mostly overlap with the solid OH03 curves and thus are not distinctly visible. This shows that our numerical integrals are converged to better than 0.01\% relative accuracy up to $L \approx 2000$, and to better than 0.03\% for $L \lesssim 3000$, and thus makes us confident that the improvement observed for OH03 over HO02 for small angular scales is not a numerical artifact. 

\section{GMV to HO02 comparison}
\label{app:gmv_ho02}
\begin{figure}
\centering
\includegraphics[width=\textwidth]{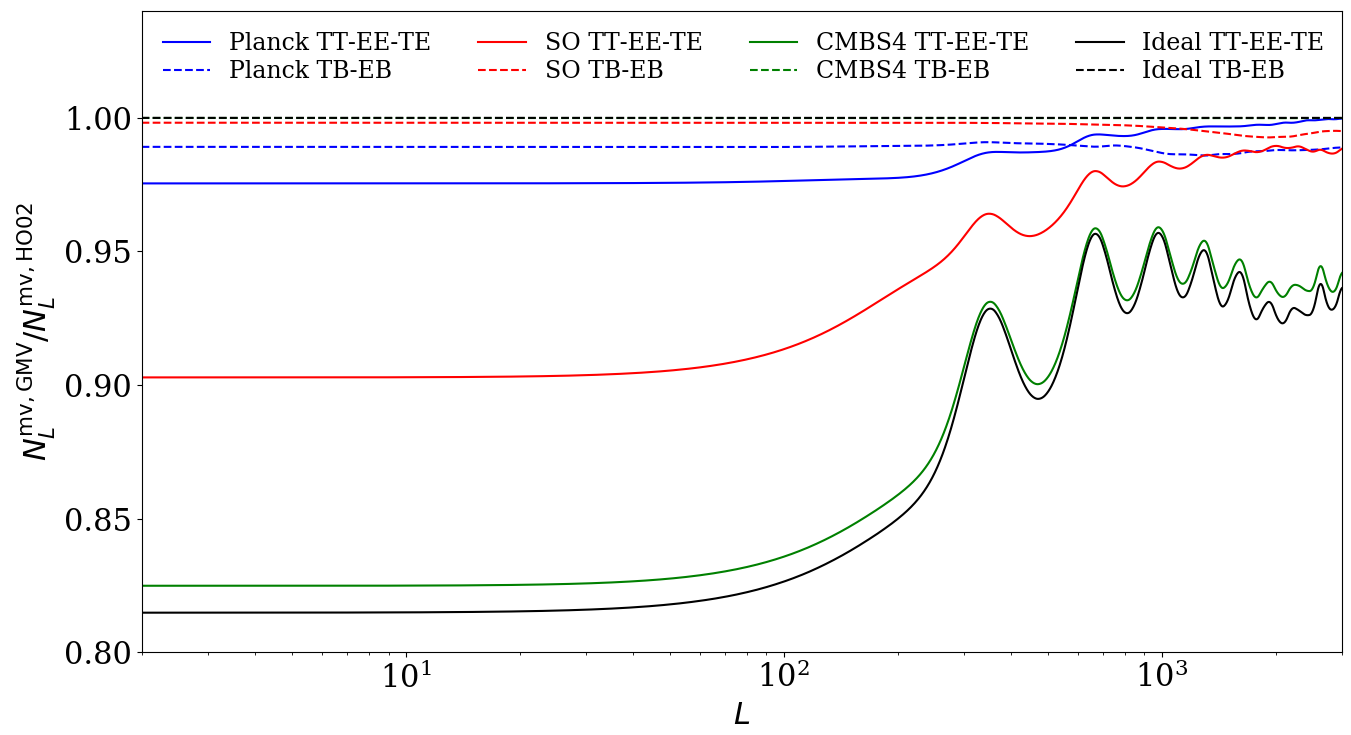}
\centering \caption{Ratio of the minimum variance reconstruction noise of the GMV and HO02 estimators for $\hat{\phi}_{\{TT, TE, EE\}}$ and $\hat{\phi}_{\{TB, EB\}}$ estimators for different experimental setups. The ideal setup corresponds to a noise-less case.}
\label{fig:GMV-HO02_comp}
\end{figure}

We perform the following exercise to compare the GMV and HO02 estimators. As mentioned at the end of Sec.~\ref{subsec:gmv}, the GMV estimator can be split into two independent estimators $\hat{\phi}^{\rm GMV}_{\{TT, TE, EE\}}$ and $\hat{\phi}^{\rm GMV}_{\{TB, EB\}}$. We compare the minimum variance reconstruction noise of these individual estimators with their HO02 counterparts i.e. $\hat{\phi}^{\rm HO02}_{\{TT, TE, EE\}}$ and $\hat{\phi}^{\rm HO02}_{\{TB, EB\}}$. This is shown in Fig.~\ref{fig:GMV-HO02_comp}, where we plot the ratio of the reconstruction noises for the GMV and HO02 versions of these two estimators. As we can see, HO02 performs almost equally well as the GMV estimator for $\{TB, EB\}$ pair. The overall improvement of the GMV estimator over HO02 estimator is thus mainly driven by $\{TT, TE, EE\}$,
especially for more sensitive SO- and CMBS4-like experiments. We also show an ideal case setup in Fig.~\ref{fig:GMV-HO02_comp} which corresponds to a noise-less experiment with the same multipole ranges as other experiments considered. HO02 estimator for a CMBS4-like setup considered here performs almost as well as the GMV estimator for $\{TB, EB\}$ pair and very slightly under-performs for the $\{TT, TE, EE\}$ set when compared to the ideal case setup.

\twocolumngrid
\bibliography{QE_lens.bib}

\end{document}